\documentclass[review]{elsarticle}

\journal{Journal of \LaTeX\ Templates}

\usepackage{graphicx}
\usepackage{float}
\usepackage{caption}
\usepackage{subfigure} 
\usepackage{url}
\usepackage{amsmath}
\usepackage{amssymb}
\usepackage[usenames,dvipsnames]{xcolor}
\usepackage{color}
\usepackage{setspace}
\usepackage{supertabular}
\usepackage{xspace}
\usepackage{soul}
\soulregister\cite7
\soulregister\ref7
\usepackage{balance}
\usepackage{amsfonts}
\usepackage[noend]{algpseudocode}
\usepackage{algorithm}
\usepackage{algorithmicx}
\usepackage{pifont}
\usepackage{epstopdf}
\usepackage{bm}
\usepackage{multicol}
\usepackage{multirow}
\usepackage{tabularx}
\usepackage{makecell}
\usepackage{array}

\newcolumntype{C}[1]{>{\vspace{0.2em}\begin{minipage}{#1}\centering\let\newline\\
\arraybackslash\hspace{0pt}}m{#1}<{\end{minipage}\vspace{0.2em}}}
\usepackage{booktabs}

\usepackage{nomencl}
\makenomenclature
\begin{document}

\captionsetup[figure]{labelfont={bf},labelformat={default},labelsep=period,name={Fig.}}

\begin{frontmatter}

\title{Energy Management Based on Multi-Agent Deep Reinforcement Learning for A Multi-Energy Industrial Park\tnoteref{mytitlenote}}
\tnotetext[mytitlenote]{This work was supported by the National Key Research and Development Program of China (Grant No.2018YFB1702300), and in part by the NSF of China (Grants No. 61731012, 62103265, 62122065 and 92167205).}


\author[mymainaddress,mysecondaryaddress,mythirdaddress]{Dafeng Zhu}

\author[mymainaddress,mysecondaryaddress,mythirdaddress]{Bo Yang\corref{mycorrespondingauthor}}
\cortext[mycorrespondingauthor]{Corresponding author}
\ead{bo.yang@sjtu.edu.cn}

\author[mymainaddress,mysecondaryaddress,mythirdaddress]{Yuxiang Liu}
\author[mymainaddress,mysecondaryaddress,mythirdaddress]{Zhaojian Wang}
\author[myfourthaddress]{Kai Ma}
\author[mymainaddress,mysecondaryaddress,mythirdaddress]{Xinping Guan}

\address[mymainaddress]{Department of Automation, Shanghai Jiao Tong University, Shanghai 200240, China}
\address[mysecondaryaddress]{Key Laboratory of System Control and Information Processing, Ministry of Education of China, Shanghai 200240, China}
\address[mythirdaddress]{Shanghai Engineering Research Center of Intelligent Control and Management, Shanghai 200240, China}
\address[myfourthaddress]{Key Laboratory of Industrial Computer Control Engineering of Hebei Province, Yanshan University, Qinhuangdao 066004, China}


\begin{abstract}
{
Owing to large industrial energy consumption, industrial production has brought a huge burden to the grid in terms of renewable energy access and power supply. Due to the coupling of multiple energy sources and the uncertainty of renewable energy and demand, centralized methods require large calculation and coordination overhead. Thus, this paper proposes a multi-energy management framework achieved by decentralized execution and centralized training for an industrial park. The energy management problem is formulated as a partially-observable Markov decision process, which is intractable by dynamic programming due to the lack of the prior knowledge of the underlying stochastic process. The objective is to minimize long-term energy costs while ensuring the demand of users. To solve this issue and improve the calculation speed, a novel multi-agent deep reinforcement learning algorithm is proposed, which contains the following key points: counterfactual baseline for facilitating contributing agents to learn better policies, soft actor-critic for improving robustness and exploring optimal solutions. A novel reward is designed by Lagrange multiplier method to ensure the capacity constraints of energy storage. In addition, considering that the increase in the number of agents leads to performance degradation due to large observation spaces, an attention mechanism is introduced to enhance the stability of policy and enable agents to focus on important energy-related information, which improves the exploration efficiency of soft actor-critic. Numerical results based on actual data verify the performance of the proposed algorithm with high scalability, indicating that the industrial park can minimize energy costs under different demands.
}

\end{abstract}

\begin{keyword}
\texttt Multi-energy management \sep industrial park \sep multi-agent \sep counterfactual baseline\sep soft actor-critic \sep attention mechanism  
\end{keyword}

\end{frontmatter}


\section{Introduction}
With the expansion of industrial production scale, energy demands have grown rapidly, which is the main driving force of industrial parks to solve the severe problems of low efficiency and high cost. To solve these issues as well as meet the multi-energy demands, energy hubs (EHs), including combined heat and power (CHP) units, boilers and energy storages, are introduced for multi-energy management in the industrial park. In EHs, multi-energy devices can be used to 
reduce energy cost \cite{Lu2020A}, optimize facility operation \cite{Jadidbonab2020Short}, and shift supply/demand \cite{Heidari2020Stochastic}. 

Many studies have been done on the multi-energy management of industrial parks. Liu et al. \cite{Liu2020Heat} establish a multi-energy framework based on Stackelberg game for an industrial park and consider bi-directional energy demand conversion to achieve peak load transfer.  Wei et al. \cite{Wei2021Distribution} propose a locational marginal price for multi-energy industrial parks to enhance the economic gains and distribute the electricity bill.
Liu et al. \cite{Liu2018Co} evaluate the benefits of eco-industrial development and demonstrate the feasibility of an accounting-based approach in an industrial park. Shahidehpour et al. \cite{Shahidehpour2021Two} propose a data-driven method to deal with the challenges of the optimal energy planning for an industrial park. However, these studies mainly concentrate on the optimization of the system to obtain solutions for energy scheduling, which usually leads to difficulties in convergence, and additional coordination overhead of energy devices is essential. To meet the real-time demand in the multi-energy coupling network, several challenges are involved: (1) Considering time-varying electricity price and stochastic demand, energy must be scheduled from the perspective of long-term metric to maximize the long-term utility. (2) Coordination overhead of energy devices needs to be reduced to a minimum. 
Therefore, a self-coordinating method in a multi-energy industrial park needs to be developed without the interaction of multiple energy devices. 


For long-term optimization, the dynamic programming is used to optimize the long-term energy cost of microgrids \cite{Zeng2019Dynamic}, which is improper for multi-energy storage systems with high uncertainty \cite{Gupta2019Optimal}. 
{In the past few years, with the rapid development of artificial intelligence (AI), deep reinforcement learning (DRL) has become the focus of attention because it successfully solves the challenging sequential decision-making problem in energy systems \cite{Li2019Energy}. DRL combines the information perception of deep learning (DL) and the decision making of reinforcement learning (RL) \cite{Kou2020Safe}. Xu et al. \cite{Xu2021Multi} use RL based differential evolution to determine the associated parameters and optimal strategy in an industrial energy system. Ebell et al. \cite{Ebell2018Reinforcement} present an algorithm based on RL to control residential power flows with a battery system and a photovoltaic system.}
Mocanu et al. \cite{Mocanu2019On} explore DRL to perform optimization schedules for energy management in buildings. Ye et al. \cite{Ye2020Local} propose a novel real-time energy management strategy for multi-energy systems based on DRL method to minimize the energy costs of end-users. 
 {Although DRL has been successfully applied in energy management \cite{Guo2022Real}}, it is still a challenge to solve the multi-energy management problem in an industrial park. Most existing DRL-based studies only take into account one energy carrier, few utilize DRL to solve multi-energy management problem. 
 {Although few studies consider multi-energy scheduling, most of these studies \cite{Sheikhi2016Demand, Weigold2021Method, Wang2021Surrogate} use single-agent DRL to manage the system energy consumption where management is carried out in a centralized framework, which is intractable under the condition of a large number of variables. } 

{To reduce computational complexity and improve robustness, multi-agent reinforcement learning (MARL) is introduced where energy devices \cite{Ebell2018Coordinated} or production resourses \cite{Roesch2019Industrial} regarded as multiple agents can perceive the environment and independently adjust their energy policies to achieve the optimal performance \cite{Ebell2019Sharing}.} Different constraints and diversity of energy are the main factors for multiple agents to make decisions in the energy management problem. Additionally, the cooperation of multiple agents can improve the calculation speed and lead to more reliable solutions \cite{Busoniu2008A}.
However, since the environment sensed by agents may become nonstationary due to changing polices of other agents, only assuming that the agents are isomorphic and selecting actions related to the maximum Q-value may cause performance degradation.
In response to this phenomenon, 
we use different input dimensions of the actor and critic networks to extend the actor-critic framework, and deploy them on energy device and energy management center to deal with nonstationary problems. The critics that approximate Q-value functions take global actions and observations as inputs, which implies the property of stationary, while energy devices make decisions based on the local observations and policies. The distributed policies of all energy devices are learned through the interaction with the environment.

 In this paper, we consider energy scheduling in an industrial park, where multi-energy devices, including energy generation, storage and conversion devices, provide energy to users. If each energy device aims at its own performance objectives under given local information, it may cause poor reward due to interference of other energy devices. Therefore, to reduce the interference and ensure the performance of each energy device, optimization metric is considered from the perspective of energy management center. In order to obtain the optimal performance based on the distributed policies of each energy device, a new policy formulation method should be designed. The stochastic renewable energy and multi-energy demand have a joint impact on system performance. Therefore, energy should be scheduled from the perspective of whole and long-term metric instead of individual and short-term performance.  
 {Different from existing research \cite{Perera2021Applications}}, the energy management problem is constructed as a partially-observable Markov decision process, which aims to develop distributed joint energy scheduling for each energy device to obtain whole performance metric. Rather than explicitly solving the problem in a time slot, the policies are learned from previous experience to minimize the long-term cost and ensure the demand of users, 
 {which implicitly solves the coupling effect and makes full use of the underlying statistical characteristics of stochastic demand and renewable energy. }
 
 In order to achieve optimal energy scheduling, the synergy among energy source, storage and load is essential, which requires the full cooperation of multiple energy devices.
Therefore, we aim in proposing a multi-agent deep reinforcement learning (MADRL) algorithm based on the counterfactual baseline and soft actor-critic (SAC), which can not only learn the implicit multi-energy devices relationship, but also combine credit assignment and attention mechanism to encourage contributing agents to learn the important information and improve their learning efficiency. 
 There are some studies on applying MADRL to solve the multi-device scheduling problems in complex cooperative scenarios \cite{Nguyen2020Deep}. MADRL is characterized by obtaining efficient and reliable solutions without establishing complex models, especially in the cooperative setting of interdependent, sequential and correlated industrial production.  
  {Different from other MADRL \cite{Lu2020Multi, Li2021A}}, the proposed algorithm not only improves  the stability of policies by integrating observable information into its own action value function estimation according to the importance of the information, but also has better scalability as the number of agents increases. The main contributions of this paper are summarized as follows.

\begin{itemize}
\item
 {To solve the nonstationary problem and reduce coordination overhead and interference of energy devices, an multi-energy management framework (MEMF) achieved by centralized training at the industrial energy management center and decentralized execution at each energy device side is proposed, where the center assists in formulating joint energy scheduling policies for each energy device. }
\item 
To reduce the calculation complexity and improve the calculation speed, a MADRL energy scheduling algorithm based on the counterfactual baseline and SAC in an industrial park is proposed, which encourages agents to learn important information by allocating reward and exploring all possible optimal paths to schedule energy devices and minimize the long-term energy cost. In addition, a novel reward is designed by Lagrange multiplier method to ensure the capacity constraints of energy storage. 
\item To avoid the inefficiency caused by the non-selective use of all information, and improve the stability of the policy and the efficiency of cooperation among energy devices/agents, an attention mechanism is introduced to enable agents to focus on important energy-related information, such as time-varying energy demand and price, which improves the exploration efficiency of SAC, instead of learning all the information in the industrial environment.
\end{itemize}

The remainder of this work is organized as follows. In Section 2, the system model of an industrial park is presented. In Section 3, a MADRL algorithm based on the attention mechanism is adopted. In Section 4, the numerical results based on actual data are shown. The paper is concluded and future research is given in Section 5.

\begin{table}
\centering
\caption{{Nomenclature}}
\begin{tabular}{m{1.4cm}|m{10cm}}
\hline
Symbol & Interpretation\\
\hline
$t$ &  time period, $t \in \{1, 2,..., T\}$\\
$k$ &  EH, $k \in \{1, 2,..., K\}$\\
$i$ &  industrial user, $i \in \{1, 2,..., I\}$\\
$j$ &  agent, $j \in \{1, 2,..., N\}$\\
$B_{k}(t)$ &  electricity of battery $k$\\
$W_{k}(t)$ &  thermal energy of hot water tank $k$\\
$C_{ke}(t)$ &  electricity charged into battery $k$\\
$C_{kh}(t)$ &  thermal energy  charged into hot water tank $k$\\
$D_{ke}(t)$ &  electricity discharged from battery $k$\\
$D_{kh}(t)$ &  thermal energy discharged from hot water tank $k$\\
$E_{kCHP}(t)$ &  electricity generation of CHP unit $k$\\
$H_{kCHP}(t)$ &  heat generation of CHP unit $k$\\
$G_{kCHP}(t)$ &  gas consumption of CHP unit $k$\\
$H_{kb}(t)$ &  heat generation of boiler $k$\\
$G_{kb}(t)$ &  gas consumption of boiler $k$\\
$E(t)$ &  electricity purchased from the electricity company\\
$G(t)$ &  gas purchased from the gas company\\
$E_{o}(t)$ &  electricity sold back to the electricity company\\
$X_{tot}(t)$ &  total available energy, $X \in \boldsymbol X= \{E, H, G\}$\\
$R(t)$ &  renewable energy generation\\
${X}_{i}(t)$ &  energy demand of user $i$\\
$p_{e}(t)$ &  electricity price\\
$p_{g}(t)$ &  gas price\\
$p_{o}(t)$ &  electricity price sold back to the electricity company\\
$\pi$ &  policy of energy scheduling\\
$r^{\pi}(t)$ &  reward of the industrial park\\
$\gamma$ &  discounted factor\\
$W^{\pi}_{\gamma}$ &  discounted function\\
$a_{B}$ &  action of battery agent\\
$a_{CHP}$ &  action of CHP agent\\
$Q$ &  action value function\\
$L(\phi^Q)$ &  regression loss of Q-network\\
$y_j$ &  target function\\
$J(\pi_{\theta})$ &  cumulative rewards\\
$\rho$ &  parameter which balances maximum entropy and rewards\\
$b(s)$ &  baseline of Q-value function\\
$z_j$ &  a weighted sum of contribution from other agents\\
$A_j(s,a)$ &  advantage function\\
\hline
\end{tabular}

\label{T1}
\end{table}

\section{System Model}
\subsection{Industrial Park}
\begin{figure}
  \centering
  \includegraphics[width=0.7\textwidth]{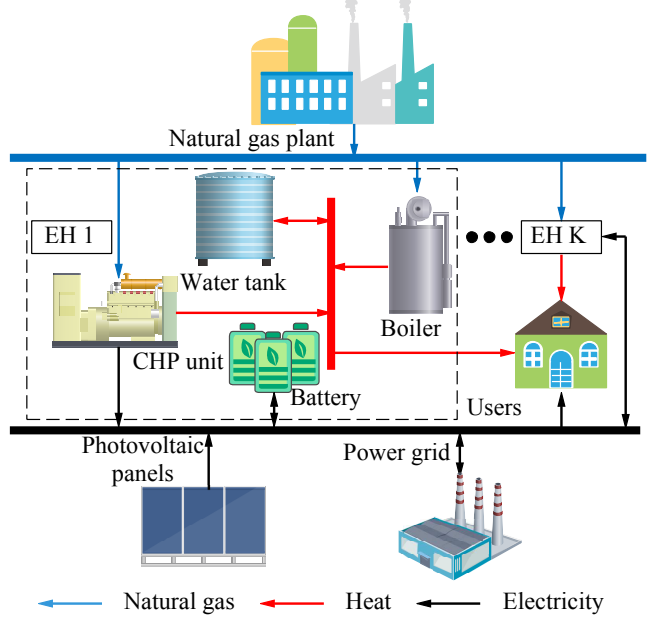}
  \caption{{Energy flows of the industrial park}}
  \label{fig1}
\end{figure}

We consider a system including electricity and gas utility companies and an industrial park
 with three types of energy: electricity, heat and gas. The park consists of users, photovoltaic panels and EHs, including CHP units, boilers, water tanks and batteries, as shown in Fig.~\ref{fig1}.
The park can harvest renewable energy generated by photovoltaic panels, and generate heat and electricity with fixed ratios by CHP units. Meanwhile, the park can store extra energy by batteries and water tanks for the demand in the future. The energy management center takes charge of the energy market of the industrial park, the operation of EHs, and the energy trading with the gas and electricity utility companies. The industrial park has $K$ EHs. The energy devices are modeled for EH $k$ in the next section. 
{For ease of reference, the nomenclature is summarized in Table~\ref{T1}.}
\subsection{Energy Hub}
The model of EH $k$, including a battery, a hot water tank, a CHP unit and a boiler, is denoted as:
\begin{equation}
B_{k}(t+1)=B_{k}(t)+\eta_{cke}C_{ke}(t)-\frac{1}{\eta_{dke}}D_{ke}(t)
\label{A1}
\end{equation}
\begin{equation}
W_{k}(t+1)=W_{k}(t)+\eta_{ckh}C_{kh}(t)-\frac{1}{\eta_{dkh}}D_{kh}(t)
\label{A3}
\end{equation}
\begin{equation}
0\leq B_{k}(t) \leq B_{k,max}
\label{Bm}
\end{equation}
\begin{equation}
0\leq W_{k}(t) \leq W_{k,max}
\label{Wm}
\end{equation}
\begin{equation}
0\leq C_{ke}(t) \leq C_{ke,max}, 0\leq D_{ke}(t) \leq D_{ke,max}
\label{Cem}
\end{equation}
\begin{equation}
0\leq C_{kh}(t) \leq C_{kh,max}, 0\leq D_{kh}(t) \leq D_{kh,max}
\label{Chm}
\end{equation}
\begin{equation}
\begin{split}
E_{kCHP}(t)&=\eta_{kpg}G_{kCHP}(t)\\
H_{kCHP}(t)&=\eta_{khg}G_{kCHP}(t)
\end{split}
\label{CHP1}
\end{equation}
\begin{equation}
\begin{split}
0&\leq E_{kCHP}(t) \leq E_{kCHP,max}\\
0&\leq H_{kCHP}(t) \leq H_{kCHP,max}
\end{split}
\label{CHP2}
\end{equation}
\begin{equation}
H_{kb}(t)=\eta_{kbg}G_{kb}(t)
\label{bo1}
\end{equation}
\begin{equation}
0\leq H_{kb}(t) \leq H_{kb,max}
\label{bo2}
\end{equation}
where (\ref{A1})-(\ref{Chm}) denote the model of the electricity and heat storages. The amount $B_{k}(t+1)$ of electricity storage at time slot $t+1$ is equal to the amount $B_{k}(t)$ at time slot $t$, plus the amount $\eta_{cke}C_{ke}(t)$ of charging and minus the amount $\frac{D_{ke}(t)}{\eta_{dke}}$ of discharging. The charging source is the low-price utility power and the remaining renewable energy after serving demand, and its amount is determined by the proposed algorithm. The amount $W_{k}(t+1)$ of equivalent thermal energy storage is equal to the amount $W_{k}(t)$, plus the amount $\eta_{ckh}C_{kh}(t)$ of charging and minus the amount $\frac{D_{kh}(t)}{\eta_{dkh}}$ of discharging. (\ref{CHP1}) and (\ref{CHP2}) denote that CHP unit $k$ generates electricity $E_{kCHP}(t)$ and heat $H_{kCHP}(t)$ by consuming gas $G_{kCHP}(t)$ simultaneously, and the efficiencies are $\eta_{kpg}$ and $\eta_{khg}$, respectively. (\ref{bo1}) and (\ref{bo2}) denote that boiler $k$ generates heat $H_{kb}(t)$ by consuming gas $G_{kb}(t)$, and the efficiency is $\eta_{kbg}$.

\subsection{Energy Trading with Utility Companies}
The industrial park purchases electricity $E(t)$ at price $p_e(t)$ and natural gas $G(t)$ at price $p_g(t)$ from the electricity and gas utility companies, respectively. In addition, the industrial park can sell electricity $E_{o}(t)$ at price $p_{o}(t)$ back to the electricity company when the electricity is surplus. The constraints of energy trading with utility companies are:
\begin{equation}
\begin{split}
0&\leq E(t) \leq E_{max} \\
0&\leq G(t) \leq G_{max} \\
0&\leq E_{o}(t) \leq E_{o,max}
\label{etu}
\end{split}
\end{equation}

\subsection{Energy Balance}
The total available energy for industrial users depends on the energy flows of EHs and utility companies. 
\begin{equation}
\begin{split}
E_{tot}(t)&=\sum_{k=1}^{K}[E_{kCHP}(t)+D_{ke}(t)-C_{ke}(t)]\\&+R(t)+E(t)-E_{o}(t)\\
 G_{tot}(t)&=G(t)-\sum_{k=1}^{K}[G_{kCHP}(t)+G_{kb}(t)]\\
H_{tot}(t)&=\sum_{k=1}^{K}[H_{kCHP}(t)+H_{kb}(t)+D_{kh}(t)-C_{kh}(t)]
\label{eb}
\end{split}
\end{equation}
where $R(t)$ is the renewable energy generation. $E_{tot}(t)$, $G_{tot}(t)$ and $H_{tot}(t)$ denote the total available electricity, gas and heat, respectively. The total available energy domain $X_{tot}(t)$ can be denoted as:

\begin{equation}
\sum_{i=1}^{I} X_{i}(t)\leq X_{tot}(t)
\label{eqdt}
\end{equation}
where $X_{i}(t)$ is the energy demand of user $i$ for energy $X \in \boldsymbol X$ at time slot t,  and $\boldsymbol X = \{E, G, H\}$ denotes the set of electricity, gas and heat. 



\subsection{Optimization Problem} 

In the industrial park, the reward $r^{\pi}(t)=E_{o}(t)p_{o}(t)
-E(t)p_{e}(t)-G(t)p_{g}(t)+b_1-b_2|X_{tot}(t)-\sum_{i=1}^{I}X^*_{i}(t)|$ consists of the revenue of electricity sold to the electricity utility company, the payment of purchasing electricity and gas from the utility companies, and the utility of matching the supply and target demand $X^*_{i}(t)$. $b_1$ and $b_2$ are the utility coefficients.

According to the reward of the industrial park, the optimization problem of the energy scheduling is to find a policy $\pi$ to maximize the time average energy utility $\overline{r}$:
\begin{equation}
\begin{aligned}
\max_{\boldsymbol{M}(t)} \overline{r}^{\pi}=\max_{\boldsymbol{M}(t)} \lim_{T\rightarrow\infty}\frac{1}{T} \sum_{t=1}^{T}r^{\pi}(t)\\
\end{aligned}
\label{eq30}
\end{equation}
\begin{equation}
\nonumber
\text{s.t. } (\ref{A1}) - (\ref{eqdt})
\end{equation}
where $\boldsymbol{M}(t)$=\{$D_{ke}(t)$, $C_{ke}(t)$, $D_{kh}(t)$, $C_{kh}(t)$, $E_{o}(t)$, $E(t)$, $G(t)$, $X_{tot}(t)$\}. The problem above is an average Markov decision process (MDP) problem over a long-term horizon, which is usually solved by dynamic programming (DP). However, DP can obtain the optimal strategy under the condition that the environmental information is available, which is unrealistic, especially with a long-term horizon. Additionally, the use of DP will cause a lot of computational overhead. Thus, a novel method independent of the environmental information needs to be developed.

The lack of the transition probability models of demand and state information causes the invalidation of DP. RL is an effective approach, which can achieve the effect of DP with incomplete information of the environment and less calculation \cite{Sutton1998Reinforcement}. Thus, RL is used to solve the above problem. 

In the long-term MDP, the discounted function provides a reliable way for evaluating policies. The discounted function $W_{\gamma}^{\pi}$ is set as the cumulative sum of discounted rewards with regard to a discounted factor $\gamma$, which is denote as
\begin{equation}
\begin{aligned}
W_{\gamma}^{\pi}=\lim_{T\rightarrow\infty}\mathbb{E}[\sum_{t=1}^{T}\gamma^tr^{\pi}(t)]\\
\end{aligned}
\label{eq32}
\end{equation}

$\pi^*$ is regarded as the optimal policy of discounted function, which means $W_{\gamma}^{\pi^*}\geq W_{\gamma}^{\pi}$ for all actions and states. Based on Laurent series expansion \cite{Puterman2009Markov}, the discounted function $W_{\gamma}^{\pi}$ is expanded as 
\begin{equation}
\begin{aligned}
W_{\gamma}^{\pi}=\frac{\overline{r}^{\pi}}{1-\gamma}+W^{\pi}+w^{\pi}(\gamma)\\
\end{aligned}
\label{eq33}
\end{equation}
where $W^{\pi}=\lim_{T\rightarrow\infty}\mathbb{E}[\sum_{t=1}^{T}(r^{\pi}(t)-\overline{r}^{\pi})]$ denotes the bias function with regard to policy $\pi$, and $w^{\pi}(\gamma)$ converges to 0 as $\gamma \rightarrow 1$.
\begin{equation}
\begin{aligned}
&\lim_{\gamma\rightarrow 1}(1-\gamma)(W_{\gamma}^{\pi^*}- W_{\gamma}^{\pi})\\&=\lim_{\gamma\rightarrow 1}({\overline{r}^{\pi^*}-\overline{r}^{\pi}})\\&+\lim_{\gamma\rightarrow 1}(1-\gamma)(W^{\pi^*}+w^{\pi^*}(\gamma)-W^{\pi}-w^{\pi}(\gamma))\\&=\lim_{\gamma\rightarrow 1}({\overline{r}^{\pi^*}-\overline{r}^{\pi}})
\end{aligned}
\label{eq331}
\end{equation}

Since $0<\gamma< 1$ and $W_{\gamma}^{\pi^*}\geq W_{\gamma}^{\pi}$, ${\overline{r}^{\pi^*}\geq\overline{r}^{\pi}}$ for $\gamma\rightarrow 1$, which means that the policy $\pi^*$ is optimal policy of reward when $\pi^*$ is optimal policy of discounted function for $\gamma\rightarrow 1$.

Therefore, if $\gamma\rightarrow 1$, the problem (\ref{eq30}) of reward maximization is converted into the problem of discounted function maximization
\begin{equation}
\begin{aligned}
&\max_{\boldsymbol{M}(t)} W_{\gamma}^{\pi}
\end{aligned}
\label{eq34}
\end{equation}
\begin{equation}
\nonumber
\text{s.t. } (\ref{A1}) - (\ref{eqdt})
\end{equation}

The maximization problem of discounted function can generally be solved by RL algorithm. It is straightforward to solve the long-term average reward problems in a centralized method. However, the centralized method may cause a huge overhead due to the presence of many variables. To avoid the overhead, a joint policy for each energy device needs to be executed without coordination overhead so that each energy device makes decision independently based on the local observations.

\section{MADRL Algortihm}
Various energy devices, including CHP units, boilers, batteries, water tanks and photovoltaic panels, play a key role in adjusting the level and type of energy consumption and supply, which may reduce energy cost, and improve energy efficiency and reliability. Due to multi-energy coupling, random renewable energy and demand, the energy management problem is difficult to deal with by traditional methods like DP, which require a priori information of random processes and have high computational complexity. Therefore, we adopt RL algorithm which can achieve the effect of DP without corresponding conditions and less calculation. In addition, the cooperation of multiple agents can improve the calculation speed and lead to more reliable solutions. To schedule energy devices efficiently, a novel MADRL algorithm is adopted where the energy devices are regarded as the agents. 

\subsection{Partially-Observable Markov Decision Process}
The interaction of multiple agents is constructed as a partially-observable Markov decision process (POMDP). The POMDP is defined by a 5-tuple ($S, A, T_s,\\ R, \pi$), i.e., a state set $S$, where observation sets, $O_1$, $O_2$, ..., $O_N$, denote partial observable information about the states, such as time and electricity price; action sets of $N$ agents, $A_1$, $A_2$, ..., $A_N$; a state transition function $T_s: S \times A_1 \times ... \times A_N \rightarrow S'$, which denotes the probability distribution of the next state; a reward set of agents $R$; and policy sets $\pi_1$, $\pi_2$, ..., $\pi_N$, which map observations to actions. Both transition and reward functions depend on actions and states of each agent. 
In the industrial park, multiple energy devices/agents interact with the industrial environment and learn the optimal policies to maximize the cumulative energy utility. Each agent obtains local information and takes actions at time slot $t$, and then the multi-agent system transfers from the current state to the next state and allocates corresponding rewards to agents.  
Details of state, action and reward for each agent are as follows:
\begin{itemize}
\item State:
In the industrial park, each energy device/agent has its own observation, including its energy consumption and generation. In addition to observation, the state also contains the time and energy price.
\item Action:
For battery agent, its task is to control the charging or discharging state of the battery, which should comply with the battery restrictions mentioned above. The actions of battery agent consist of three types: discharging, idle and charging, which are between -1 and 1. 
{The actions are equally divided into 20 steps as follows: }
\begin{equation}
\begin{aligned}
a_{B}=\left\{\begin{array}{cc}
1, ..., 0.3, 0.2, 0.1 & \mbox{Charging}\\
0 & \mbox{Idle}\\
-0.1, -0.2, ..., -1 & \mbox{Discharging} \end{array}\right.
\end{aligned}
\end{equation}
where the action $a_j=1$ means energy device $j$ is working at full capacity, and $a_j=0$ means energy device $j$ is idle, and negative and positive values denote discharging and charging, respectively. The water tank agent has the same action sets as the battery agent. The actions of CHP agent are between 0 and 1, which are equally divided into 10 steps  
$a_{CHP}=0, 0.1, 0.2, ..., 1$.
The boiler agent has the same action sets as the CHP agent. 

\item Reward:
According to the reward function given in Section 2.5, the reward of the industrial park is expressed as $r(t)=E_{o}(t)p_{o}(t) 
-E(t)p_{e}(t)-G(t)p_{g}(t)+b_1-b_2|X_{tot}(t)-\sum_{i=1}^{I}X^*_{i}(t)|$. In this paper, each energy device/agent is assumed to receive a same reward $r_j(t)=r(t)$. Therefore, all agents will aim to maximize the common reward, which can be achieved by the proposed algorithm in the following section.
\end{itemize}

{Although the constraints of energy charging/discharging of batteries (\ref{Cem}) and water tanks (\ref{Chm}), heat and electricity generation of CHPs (\ref{CHP2}) and heat generation (\ref{bo2}) of boilers are satisfied by setting the maximum action value of agents}, traditional RL cannot guarantee the capacity constraints of batteries (\ref{Bm}) and water tanks (\ref{Wm}). To guide the policy toward a solution of constraint satisfying, we add a penalty term to the reward function of the battery agent $k$. The reward is revised as
\begin{equation}
{r}_{\overline{B}_k}(t)=r_{B_k}(t)-\lambda_{B_k}(t)(B_k(t)-B_{k,max})
\label{loss1}
\end{equation}

The penalized $Q$ function is denoted as 
\begin{equation}
\begin{aligned}
&{Q}_{\overline{B}_k}(s_t, a_t)=\mathbb{E}[\sum_{t'=t}^{\infty}\gamma^{t'-t}{r}_{\overline{B}_k}(t'+1)]\\&=\mathbb{E}[\sum_{t'=t+1}^{\infty}\gamma^{t'-t-1}(r_{B_k}(t')-\lambda_{B_k}(t')(B_k(t')-B_{k,max}))]\\&=Q_{B_k}(s_t, a_t)-\sum_{t'=t+1}^{\infty}\gamma^{t'-t-1}\lambda_{B_k}(t')\mathbb{E}[B_k(t')-B_{k,max}]
\end{aligned}
\label{loss2}
\end{equation}

The corresponding Lagrange multiplier can be updated as
\begin{equation}
\lambda_{B_k}(t+1)=\mbox{clip}(\lambda_{B_k}(t)+ \zeta_{B_k}(B_k(t)-B_{k,max}),0,1)
\label{loss3}
\end{equation}
\begin{center}
$\mbox{clip}(x,0,1)=\left\{\begin{array}{cc}
0, & \mbox{if } x<0\\
x, & \mbox{if } 0\leq x \leq 1\\
1, & \mbox{if } x>1 \end{array}\right.$
\end{center}
where $\zeta_{B_k}>0$ is the updating rate.

When $B_k(t)-B_{k,max}>0$, the reward function ${r}_{\overline{B}_k}<r_{B_k}$. When $B_k(t)-B_{k,max}\leq0$, the reward function ${r}_{\overline{B}_k}\geq r_{B_k}$. Since the battery agent aims to obtain larger reward, it will satisfy the battery capacity constraint. In addition, the existence of positive and negative rewards can improve the learning efficiency of the agent. When $\lambda_{B_k}(t) \rightarrow 0$, the learned policy can guarantee the capacity constraint of the battery, and ${Q}_{\overline{B}_k}(s_t, a_t)$ converges to $Q_{B_k}(s_t, a_t)$. The water tank agent adopts the same method as the battery agent to satisfy the capacity constraint of the water tank.

\subsection{DRL Framework}
The existence of multi-energy coupling indicates that the action taken by one energy device will affect the performance of other energy devices. As a result, the environments observed partially by agents become nonstationary. However, the traditional reinforcement learning (RL) is generally suited in a stationary environment. Thus, the policy cannot be learned completely independently. To solve this issue, one way is to acquire all actions and states of agents, but it will cause a huge overhead. Therefore, both completely decentralized and centralized methods are impractical. 
 {To balance the performance and overhead, a SAC model in discrete domain \cite{Wu2021Caching, Wang2020Safe}, which can improve exploration ability and robustness, is introduced to achieve centralized training by the critic network and decentralized execution by the actor network. }


During the centralized training, the interaction among the agents and the energy management center is carried out. The center obtains the states, actions, rewards and other information from each agent and learns the implicit relationship of multi-agent. The gradients of $Q$ about the joint actions are calculated on the center and sent to the agents. During the decentralized execution, each agent  only needs local information and gradients to update the policy $\pi$, and then acts respectively according to the established policies. In this way, the proposed algorithm can reduce the complexity without coordinating different agents. In order to ensure the convergence of the results, the experience replay is adopted, which uses a small size of random sample to obtain the gradients and the loss of the value $Q$. 


\subsection{Centralized Training}
The action value function $Q_{\phi}(s_t, a_t)=\mathbb{E}[\sum_{t'=t}^{\infty}\gamma^{t'-t}r_{t'}(s_{t'}, a_{t'})]$ denotes the cumulative rewards after each agent takes actions. Therefore, $Q$ obtains the implicit relationship among agents, i.e., the interaction among energy devices in the industrial park. The energy management center receives transition function $T_s$ from all agents, and stores these information in the replay buffer $D$. For simplicity, the time subscript $t$ of variables is removed. The regression loss of Q-network is denoted by
\begin{equation}
\begin{aligned}
L(\phi^Q)&=\sum_{j=1}^N\mathbb{E}_D[Q_{\phi_j}(s, a)-y_j]^2\\
y_j&=r_j + \gamma \mathbb{E}_{\pi}[Q_{\phi'_j}(s', a')]
\end{aligned}
\label{loss}
\end{equation}
where $y_j$ is the target function. $\gamma$ is the discount factor which balances the long-term reward and immediate reward, and $s'$ and $a'$ denote the next observation state and action, respectively. 

In the policy gradient method \cite{Sutton2000Policy}, agent $j$ selects a policy to maximize its expected cumulative rewards $J(\pi_{\theta})=\mathbb{E}_{D,\pi}[Q_{\phi}(s, a)]$, and its gradient is denoted as
\begin{equation}
\begin{aligned}
\nabla_{\theta_j}J(\pi_{\theta})=\mathbb{E}_{D, \pi}[\nabla_{\theta_j}\log(\pi_{\theta_j}(a_j|s_j))Q_{\phi_j}(s, a)]
\end{aligned}
\label{gra}
\end{equation}

{To explore more useful information and be more robust, the maximum entropy \cite{Haarnoja2018Soft} is introduced to adjust the policy gradient method:}
\begin{equation}
\begin{aligned}
\nabla_{\theta_j}J(\pi_{\theta})&=\mathbb{E}_{D, \pi}[\nabla_{\theta_j}\log(\pi_{\theta_j}(a_j|s_j))(-\rho \log(\pi_{\theta_j}(a_j|s_j))\\&+Q_{\phi_j}(s, a)-b(s))]
\end{aligned}
\label{rgra}
\end{equation}
where  $\rho$ is the parameter which balances maximum entropy and rewards, 
{and $b(s)$ is a baseline of the Q-value function, which will be introduced in detail below.} Meanwhile, the target function is also adjusted as
\begin{equation}
\begin{aligned}
y_j&=r_j + \gamma \mathbb{E}_{ \pi}[Q_{\phi'_j}(s', a')-\rho \log(\pi_{\theta'_j}(a'_j|s'_j)]
\end{aligned}
\label{tara}
\end{equation}

To reduce the difficulty and computational complexity of selecting a baseline, the current action value function is used to solve the marginal distribution of the current policies, which will be elaborated in the following section. 

\subsection{Attention Mechanism}
\begin{figure}
  \centering
  \includegraphics[width=0.4\textwidth]{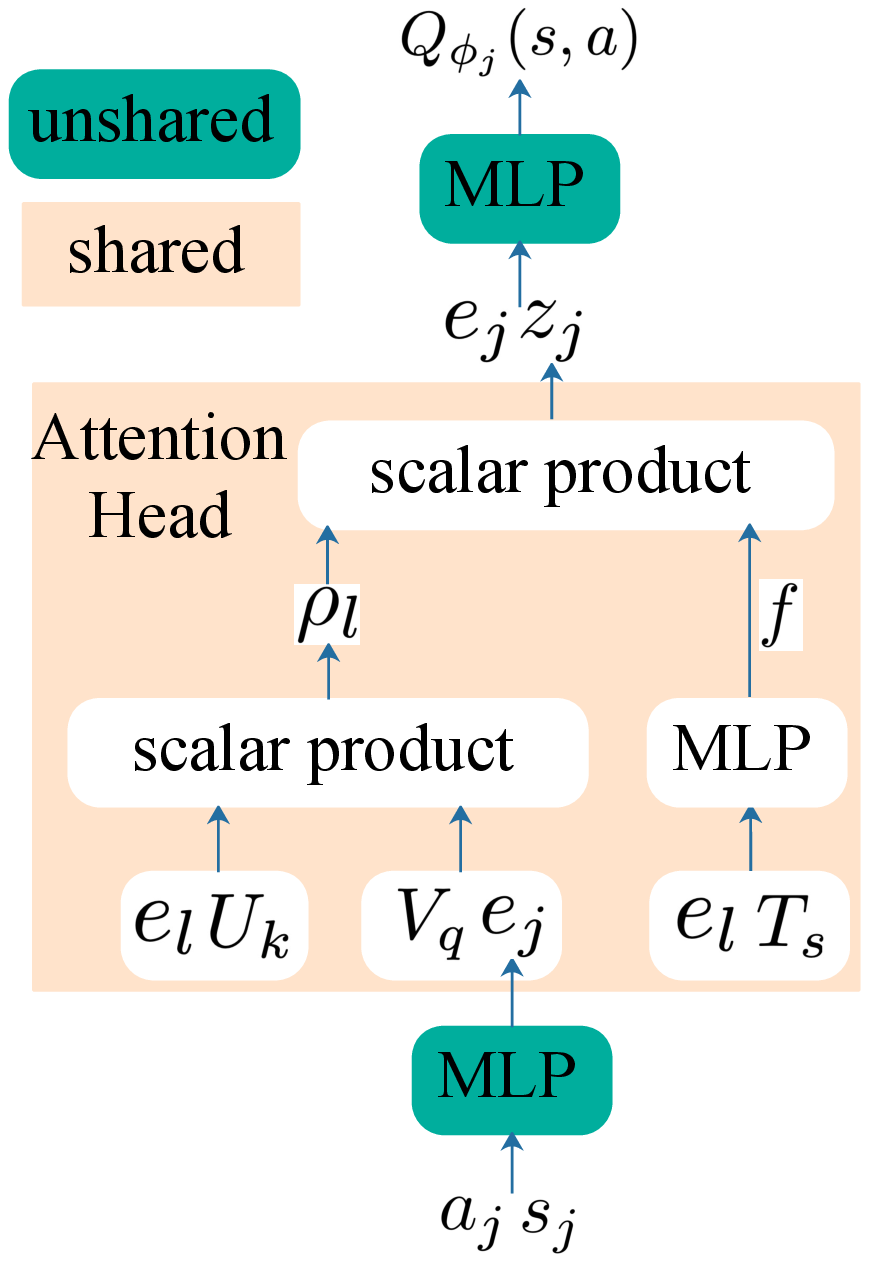}
  \caption{{The calculation process of $Q_{\phi_j}(s, a)$ for agent $j$ based on attention mechanism}}
  \label{fig33}
\end{figure}

The attention mechanism is introduced to improve the learning efficiency of the agents and the stability of the policies. The key point of the attention mechanism is that each agent can selectively focus on information that is more conducive to obtaining greater rewards when learning the critic. It is suitable for centralized training and decentralized execution, as shown in Fig.~\ref{fig33}. 

The critic receives all actions $a=(a_1, ..., a_N)$ and all states $s=(s_1, ..., s_N)$ of agents to calculate $Q_{\phi_j}(s, a)$ for agent $j$. $Q_{\phi_j}(s, a)$ is a function about the state and action of agent $j$, and the contribution $z_j$:
\begin{equation}
\begin{aligned}
Q_{\phi_j}(s, a)=g_j(h_j(s_j, a_j), z_j)
\end{aligned}
\label{qoa}
\end{equation}
where $g_j$ and $h_j$ denote the multi-layer perceptron (MLP) function. $z_j$ denotes a weighted sum of contribution from other agents:
\begin{equation}
\begin{aligned}
z_j=\sum_{l\neq j}\rho_{l}F_l=\sum_{l\neq j}\rho_{l}f(T_sh_l(s_l, a_l))
\end{aligned}
\label{cont}
\end{equation}
where $F_l$ denotes a one-layer embedding function of agent $l$ transformed by shared matrix $T_s$, and $f$ denotes a leaky rectified linear unit (ReLU). The attention mechanism can be regarded as a key-value model where agents query the information to estimate the value function \cite{Oh2016Control}. 
{As the name implies, the essence of the attention mechanism is to screen the input information to retain valuable information and filter out unimportant information. In mathematical language, its expression is the attention weight, which is multiplied by the input vector to get the filtered information. In the attention mechanism, the target to be filtered is the encoder output, and the input source for generating the attention weight is the hidden layer output from the decoder.  In general, a bilinear mapping is used to calculate attention weight. And the bilinear matrix is the parameter matrix of the fully connected layer that needs to be learned. In order to reduce the rank of the bilinear matrix, the product of the low-rank matrices $U_k$ and $V_q$ is used to obtain attention weight $\rho_l$ by fusing $e_j=h_j(s_j, a_j)$ and $e_l$  \cite{Kim2016Hadamard}: }
\begin{equation}
\begin{aligned}
\rho_l=p_l\exp(e_l^{T}U_k^{T}V_{q}e_j)
\end{aligned}
\label{awei}
\end{equation}
{where $p_l$ is the linear coefficient; $e_j$ and $e_l$ are transformed to a query value and a key value, respectively. The match between query value and key value is adjusted according to the ranks of the $V_q$ and $U_k$ to prevent the gradient from disappearing \cite{Vaswani2017Attention}. 

In the experimental design, several attention heads are used. Each head uses a set of independent parameters ($U_k$, $V_q$, $T_s$), which generate the contribution of other agents to agent $j$, and  contributions from all attention heads are connected as a vector. It is worth mentioning that each head can focus on the weighted contribution of other agents from a different perspective.}
In addition, the weight's feature extractors, keys and values are shared among multiple agents, so that the features of all agents are in the same space after network processing and transformation. It is feasible to share parameters between different agents no matter in a cooperative environment or a competitive environment, because the approximation of the action value function in a multi-agent system is essentially a multi-objective regression problem. This parameter sharing mechanism allows each agent to 
learn effectively in complex environments. Just by adding more encoders during training, this approach can be extended to include other information, such as global state, rather than just local actions and observations. 

Owing to the parameter sharing, critics are updated together to minimize the joint loss function:
\begin{equation}
\begin{aligned}
\phi^Q&\leftarrow\phi^Q-\alpha\nabla L(\phi^Q)\\
L(\phi^Q)&=\sum_{j=1}^N\mathbb{E}_{ D}[Q_{\phi_{j}}(s, a)-y_j]^2\\
y_j&=r_j + \gamma \mathbb{E}_{\pi}[Q_{\phi'_{j}}(s', a')-\rho \log(\pi_{\theta'_j}(a'_j|s'_j))]
\end{aligned}
\label{sloss}
\end{equation}
where $\alpha$ denotes the update stepsize, and $\phi'$ and $\theta'$ denote the network parameters of the target critics and actors, respectively. $Q_{\phi'_j}$ is the action-value estimate of agent $j$ based on states and actions of all agents. 

After introducing attention mechanism, the actors are updated as:
\begin{equation}
\begin{aligned}
\theta_j&\leftarrow\theta_j+\beta \nabla_{\theta_j}J(\pi_{\theta})\\
\nabla_{\theta_j}J(\pi_{\theta})&=\mathbb{E}_{D, \pi}[\nabla_{\theta_j}\log(\pi_{\theta_j}(a_j|s_j))(Q_{\phi_{j}}(s, a)\\&-\rho \log(\pi_{\theta_j}(a_j|s_j))-b(s, a_K))]
\end{aligned}
\label{agra}
\end{equation}
where $K$ represents the set of all other agents except $j$. 
{To evaluate the contribution of specific action, an advantage function which uses a baseline is introduced. The baseline is obtained by solving the marginal distribution problem of the current agent's strategy, which can avoid designing additional default actions and reduce simulation calculations.} In addition, the advantage function can solve the credit allocation problem in the multi-agent environment, i.e., assign rewards to encourage those agents that are more helpful to the entire multi-agent task, and then promote them to learn excellent strategies. The advantage function is expressed as:
\begin{equation}
\begin{aligned}
A_{j}(s, a)&=Q_{\phi_{j}}(s, a)-b(s, a_K)\\
b(s, a_K)&=\mathbb{E}_{\pi}[Q_{\phi_j}(s, (a_j, a_K))]
\end{aligned}
\label{adva}
\end{equation}

Different from general advantage function which requires a global reward and same action space, the proposed advantage function based on the attention mechanism can achieve a more flexible and general form of baseline without these requirements. This makes the weighted sum of encodings $z_j$ for other agents and the decomposition of encoding $e_j$ for agent $j$ more simple. 
{For discrete policies, the expected return $Q_j(s, (a_j, a_K))$ for each possible action that agent $j$ might take can be used to calculate the baseline. Its expectation is denoted by:}
\begin{equation}
\begin{aligned}
\mathbb{E}_{\pi}[Q_{\phi_{j}}(s, (a_j, a_K))]=\sum_{a'_j \in A_j}\pi(a'_j|s_j)Q_{j}(s, (a'_j, a_K))
\end{aligned}
\label{exre}
\end{equation}
{where $a_j$ needs to be removed from $Q_j$, and a $Q_j$ value is output for each action. The observation encoding $e_j=h_{o_j}(o_j)$ is added for each agent to replace $e_j=h_j(s_j, a_j)$ mentioned above, and function $g$ is adjusted to output values for all possible actions instead of outputting a value for the input action.} For continuous policies, the above expected return can be estimated by learning a value head or sampling from the policy of agent $j$. The implementation process of MADRL algorithm is shown in Algorithm 1.

\begin{algorithm}[h]
\floatname{algorithm}{Algorithm}
\footnotesize
\caption{: MADRL Algorithm based on Counterfactual Baseline and SAC}
\label{alg1}
\begin{algorithmic}[1]
  \State Initialize critic and actor networks, the target network parameters of each agent, and replay buffer $D$.  
    \For{episode=1, ..., $E$ }
    \State Initialize observation states $s$
     \For{$ t=1, ..., T$}
\State  Select and execute actions $a$
\State  Acquire rewards $r$ and observation states $s'$
\State  Calculate $Q_{\phi}(s, a)$ according to Fig.~\ref{fig33}.
\State Store transitions in replay buffer $D$
\For{agent $j$}
\State Sample random samples from $D$
\State Set $y_j=r_j + \gamma \mathbb{E}_{\pi}[Q_{\phi'_{j}}(s', a')-\rho \log(\pi_{\theta'_j}(a'_j|s'_j))]$
\State Update critic by minimizing the joint loss function:
$L(\phi^Q)=\sum_{j=1}^N\mathbb{E}_{ D}[Q_{\phi_{j}}(s, a)-y_j]^2$
\State Update actor using the policy gradient:
$\nabla_{\theta_j}J(\pi_{\theta})=\mathbb{E}_{D, \pi}[\nabla_{\theta_j}\log(\pi_{\theta_j}(a_j|s_j))(Q_{\phi_{j}}(s, a)-\rho \log(\pi_{\theta_j}(a_j|s_j))-b(s, a_K))]$
\EndFor
\State \ \textbf{end for}
\EndFor 
\EndFor 
\State \ \textbf{end for}
\State \ \textbf{end for} 
\label{code:recentEnd}
\end{algorithmic}
\end{algorithm}

\section{Simulation}
In this section, the simulation based on the real data is conducted to evaluate the performance of the proposed algorithm using PyTorch-Gym framework, which combines the PyTorch tensor library with the OpenAI Gym architecture. 
\subsection{Setup}
\begin{figure}
\centering
\begin{minipage}{1\linewidth}
  \centerline{\includegraphics[width=0.7\textwidth]{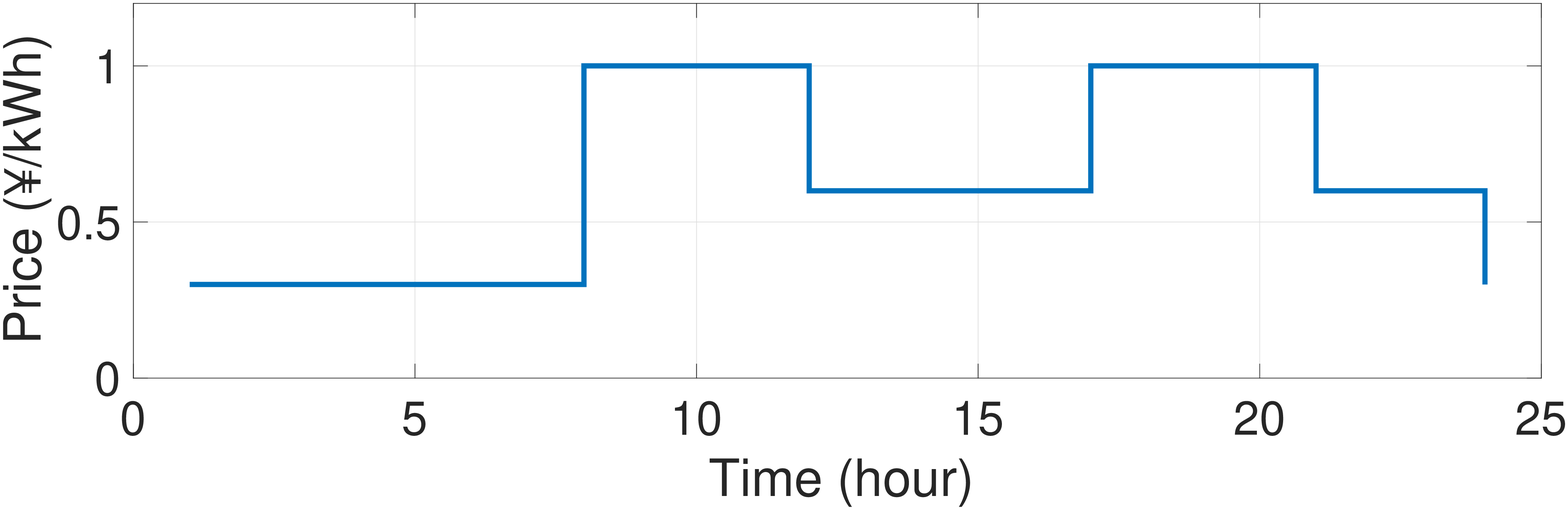}}
  \centerline{\scriptsize{(a) Electricity price}}
\end{minipage}
\begin{minipage}{\linewidth}
  \centerline{\includegraphics[width=0.7\textwidth]{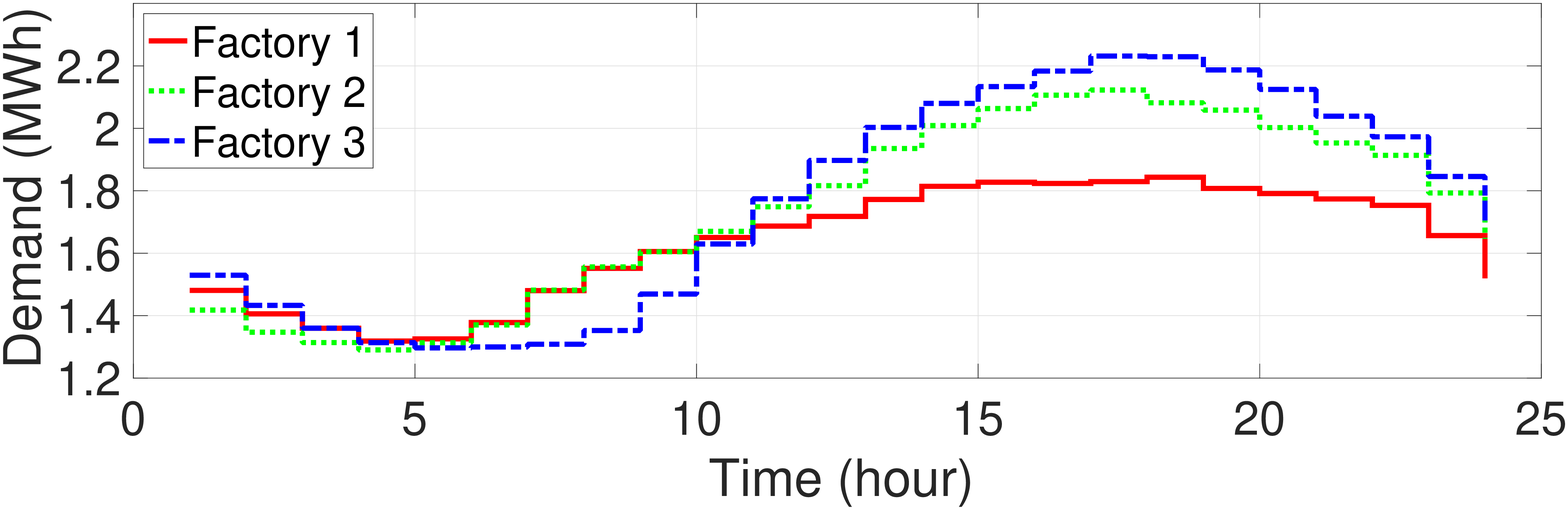}}
  \centerline{\scriptsize{(b) Electricity demand}}
\end{minipage}
\caption{{Energy data. }}
\label{fig2}
\end{figure}

For simplicity, we consider an industrial park consisting of one EH and three factories. Each EH has a CHP unit, a water tank, a battery and a boiler. Therefore, the number of agents is $N=4$. Then, the scalability of the proposed algorithm will be verified by increasing the number of agents/EHs. For different EHs, the coefficients of same kinds of energy devices are set to be same.
The parameters of efficiency are $\eta_{cke}=\eta_{dke}=\eta_{ckh}=\eta_{dkh}=98\%$, $\eta_{kpg}=\eta_{khg}=35\%$, $\eta_{kbg}=80\%$, respectively. The reward coefficients are $b_1=20, b_2=2$. Other parameters are summarized as follows: 
 $p_{g}(t)=0.3$ yuan/kWh, $B_{k,max}=4$MWh, $C_{ke,max}=D_{ke,max}=1$MWh, $W_{k,max}=4$MWh, $C_{kh,max}=D_{kh,max}=1$MWh.
The price provided by the State Grid Jiangsu Electric Power Co., Ltd  \cite{Illinois} is shown in Fig.~\ref{fig2}(a). The target electricity load provided by PJM \cite{PJM} is given in Fig.~\ref{fig2}(b). The data of photovoltaic systems provided by Renewables.ninja \cite{Institute} is shown in Fig.~\ref{fig6}(a). 


For the training of all experiments, the discount factor $\gamma$ is set as 0.95. The parameter $\rho$ which balances maximum entropy and rewards is 0.01, and 4 attention heads are used in the attention mechanism. In addition, we use 32 samples for the minibatch of random sample from replay buffer $D$, and the size of $D$ is 1000. The learning parameters of simulation are shown in Table~\ref{TAB01}.

\begin{table}
\small
\centering
\caption{Learning Parameters}
\begin{tabular}{l  l }
\hline
Parameters& Values  \\
\hline
Discount factor ($\gamma$) & 0.95  \\
Balance parameter ($\rho$)& 0.01  \\
Number of attention heads & 4 \\
Size of replay buffer $D$ & 1000 \\
Size of random samples &32   \\
\hline
\end{tabular}
\label{TAB01}
\end{table}

To verify the effectiveness of the proposed algorithm, several methods are used to compare as follows: deep deterministic policy gradient algorithm (DDPG) \cite{DDPG} and multi-agent deep deterministic policy gradient algorithm (MADDPG) \cite{MADDPG}. 
A restricted version of the proposed algorithm is also used for comparison, and the attention weight of the version is fixed to be $1/(N-1)$, which causes the model fail to focus on specific agents. All methods above are carried out with the same condition. The parameters of each algorithm are adjusted according to performance, and they remain unchanged for that algorithm. The comparison of different methods is shown in Table~\ref{TAB001}.

\begin{table}
\small
\centering
\caption{Different Methods}
\begin{tabular}{l l l }
\hline
 Methods& Base & Interaction \\
\hline
DDPG & DDPG & Observation  \\

MADDPG & DDPG & Observation  \\

Restricted Version & SAC & Restricted Attention  \\

Proposed Algorithm & SAC & Attention  \\
\hline
\end{tabular}
\label{TAB001}
\end{table}

\subsection{Performance Verification}

\begin{figure*}
\centering
\begin{minipage}{1\linewidth}
  \centerline{\includegraphics[width=\textwidth]{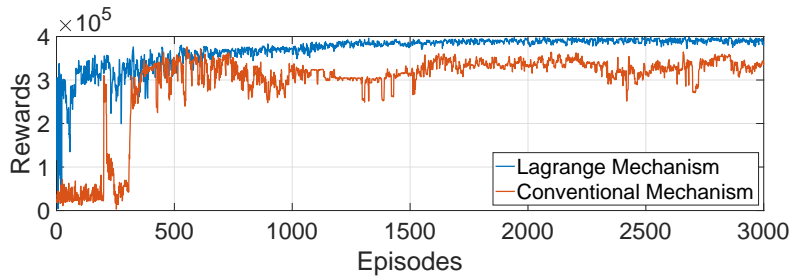}}
\end{minipage}
\caption{{Total rewards under revised reward and traditional reward.}}
\label{fig14}
\end{figure*}

 {To verify the performance of the reward designed by Lagrange mechanism, the comparison under conventional reward mechanism and Lagrange reward mechanism is given. During training, the battery charging often exceeds the capacity limit under the conventional reward mechanism. In order to obtain the total rewards under the conventional mechanism, we give a great penalty to prevent the electricity of batteries from exceeding the capacity limit. The comparison of total rewards under the conventional mechanism and Lagrange mechanism is given in Fig.~\ref{fig14}. The total rewards under the Lagrange mechanism is higher than the one under the conventional mechanism, and the learning efficiency under the Lagrange mechanism is better than the one under the conventional mechanism. Therefore, the Lagrange reward mechanism can improve the learning efficiency and ensure that the battery capacity meets the constraints.}

\begin{figure*}
\centering
\begin{minipage}{1\linewidth}
  \centerline{\includegraphics[width=\textwidth]{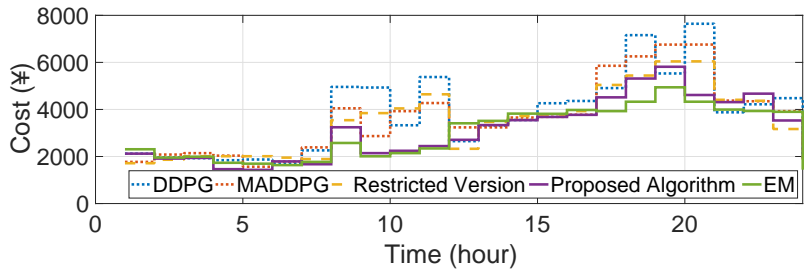}}
\end{minipage}
\caption{{Costs across 24 time slots under different methods.}}
\label{fig3}
\end{figure*}

{In order to verify the performance of the proposed algorithm, assuming that all information is known, the optimal solution can be directly calculated by exhaustive method (EM). Fig.~\ref{fig3} shows the costs across 24 time slots implemented by different methods in the industrial environment, and Table~\ref{TAB2} shows the total cost implemented by different methods. According to the figure and table, the cost of proposed algorithm is lower than the costs of other methods. The optimal solution has the same trend with proposed algorithm. The total cost can continuously approach the theoretical optimal cost after continuous iteration. However, since each agent in proposed algorithm cannot know global data, the total cost is about 3.7\% different from the theoretical optimal cost.} 

\begin{table}
\small
\centering
\caption{{Total cost implemented by different methods}}
\begin{tabular}{l l }
\hline
 Methods & Total cost (¥) \\
\hline
DDPG & 90240 \\

MADDPG & 86908  \\

Restricted Version & 83138  \\

Proposed Algorithm & 74133  \\
EM & 71493 \\
\hline
\end{tabular}
\label{TAB2}
\end{table}

Since DDPG easily causes overestimation of $Q$ and can not explore state action space well, the total cost under DDPG is high. MADDPG transfers the state transition information of all agents to the critic of each agent, and the observation spaces in the environment are relatively large for agents, so the total cost under MADDPG is high. Due to the lack of the attention mechanism, the restricted version also causes a high cost. Fig.~\ref{fig3} denotes that the proposed algorithm achieves lower costs in most cases. At the low-price time, the costs under the proposed algorithm may be higher because the park purchases low-price electricity to charge into battery instead of consuming electricity discharged by battery. When the electricity price is high, it can discharge the battery instead of purchasing much high-price electricity. Thus, the total cost of the proposed algorithm is lower. It is worth noting that the cost is high and fluctuates greatly under DDPG at high-price time, which causes the cost of DDPG to be lower than the cost of the proposed algorithm at 19:00. 

\begin{figure*}
\centering
\begin{minipage}{1\linewidth}
  \centerline{\includegraphics[width=\textwidth]{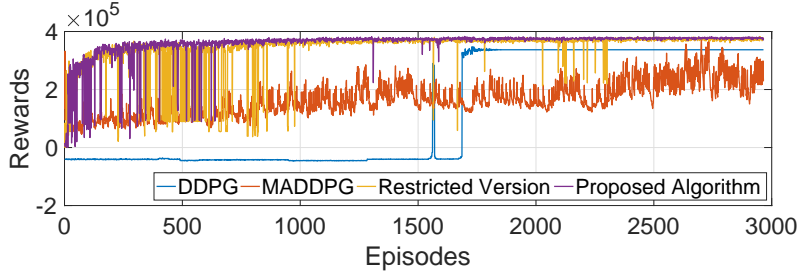}}
\end{minipage}
\caption{{Rewards implemented by different methods.}}
\label{fig4}
\end{figure*}

\begin{figure}
\centering
\begin{minipage}{1\linewidth}
  \centerline{\includegraphics[width=0.7\textwidth]{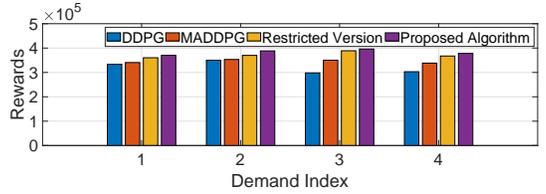}}
  \centerline{\scriptsize{(a) Comparison of the rewards under different demands}}
\end{minipage}
\begin{minipage}{1\linewidth}
  \centerline{\includegraphics[width=0.7\textwidth]{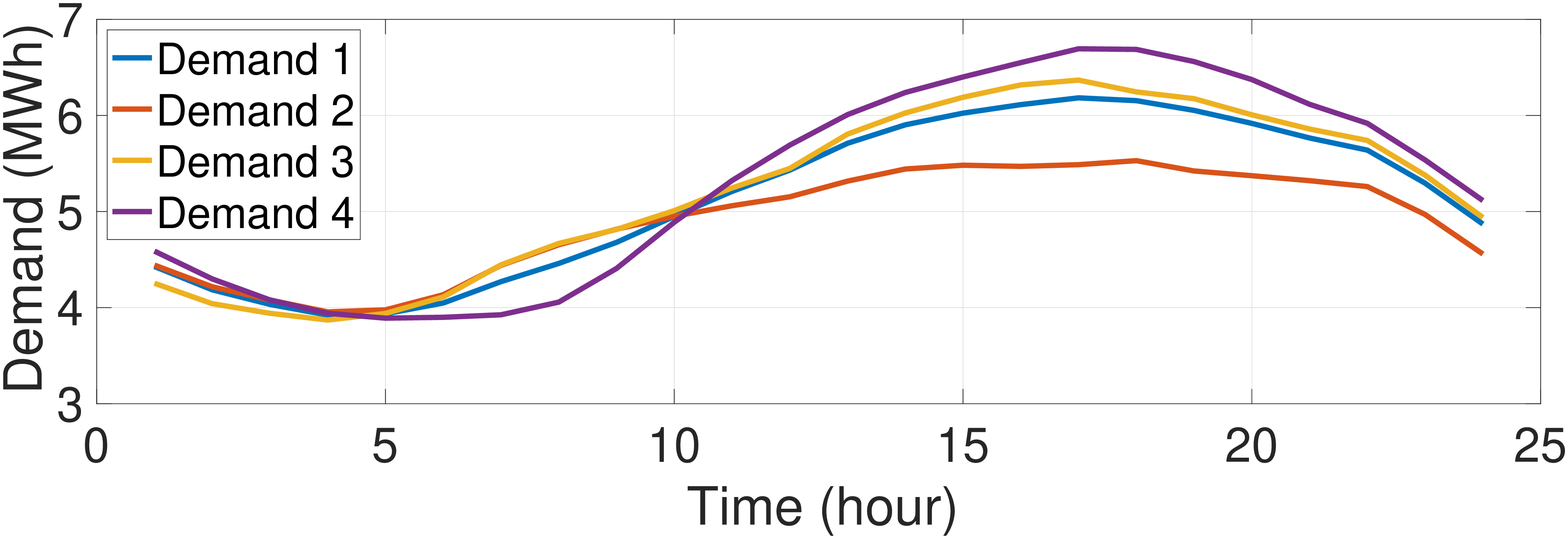}}
  \centerline{\scriptsize{(b) Different demands }}
\end{minipage}
\caption{{Rewards implemented by different methods under different demands.}}
\label{fig40}
\end{figure}

Fig.~\ref{fig4} shows the convergence situation implemented by different methods. The proposed algorithm achieves faster convergence than the restricted version owing to the attention mechanism. The proposed algorithm converges in about 1600 iterations, while the restricted version converges in about 2300 iterations. Although DDPG achieves fast convergence, its reward is lower.
{ The reward of the proposed algorithm is greater than those of other methods.} In addition, considering the stochastic of the demand, the comparison of the rewards under different demands is given in Fig.~\ref{fig40}(a), which verifies the performance of the proposed algorithm when the demand changes in Fig.~\ref{fig40}(b). 

\begin{figure}
\centering
\begin{minipage}{1\linewidth}
  \centerline{\includegraphics[width=0.7\textwidth]{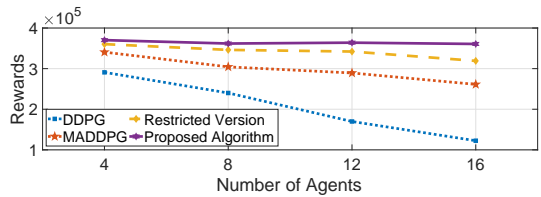}}
\end{minipage}
\caption{{Rewards implemented by different methods as the number of agents increases.}} 
\label{fig5}
\end{figure}

Then, Fig.~\ref{fig5} shows a comparison of the rewards obtained by the proposed algorithm and other methods as the number of agents increases. There is the same trend when the reward coefficients change. When the number of agents increases, it means that more energy can be scheduled, so the demand of users will also increase proportionally to make full use of these energy devices. Although MADDPG performs well on the version of 4 agents, the performance cannot be maintained as agents are added. The rewards of the proposed algorithm remain stable as the number of agents increases. Table~\ref{TAB02} shows that the improvement of the proposed algorithm grows with the number of agents over other methods. The reason is that the proposed algorithm can focus on information which need more attention by the attention mechanism, unlike MADDPG which uses all information non-selectively. Therefore, the performance of the proposed algorithm is better as the number of agents increases. 

\begin{table}
\small
\centering
\caption{The improvement of proposed algorithm over different methods as agents are added}
\begin{tabular}{l l l l l}
\hline
Number of Agents & 4 & 8 & 12 & 16 \\
\hline
Over DDPG (\%) & 27 & 51 & 114 & 194 \\

Over MADDPG (\%) & 8.7 & 19 & 26 & 38  \\

 Over Restricted Version (\%) & 2.7 & 4.5 & 6.4 & 13  \\
\hline
\end{tabular}
\label{TAB02}
\end{table}

\begin{figure}
\centering
\begin{minipage}{1\linewidth}
  \centerline{\includegraphics[width=0.7\textwidth]{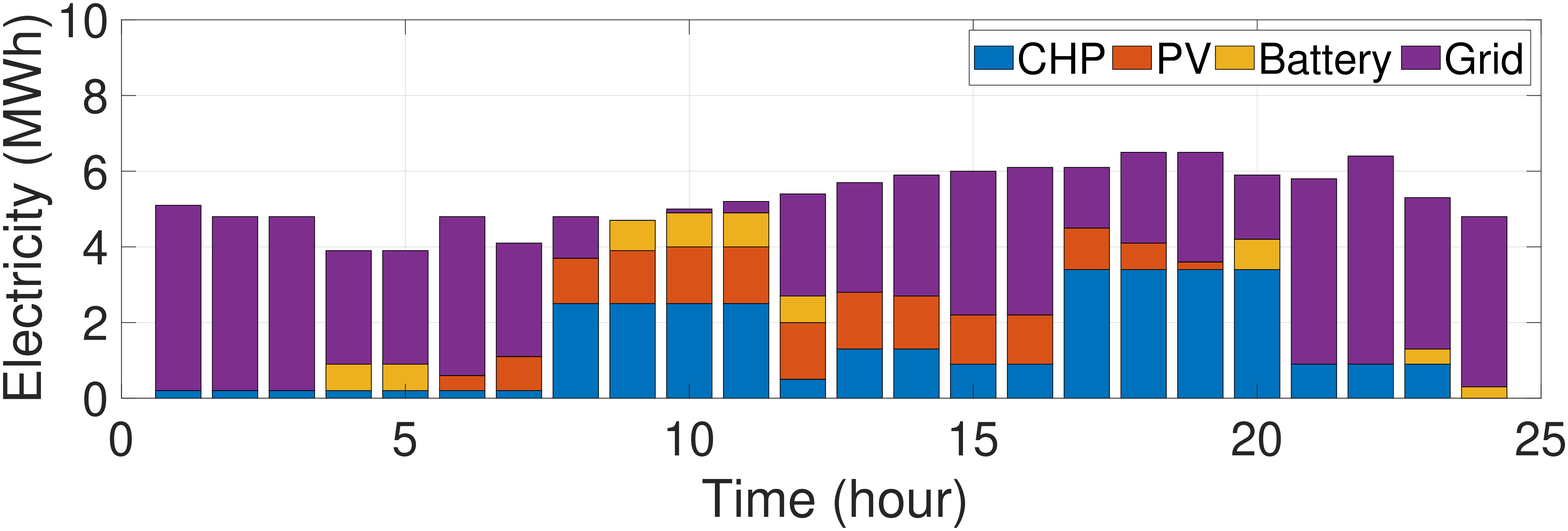}}
  \centerline{\scriptsize{(a) Electricity generation/discharging}}
\end{minipage}
\begin{minipage}{1\linewidth}
  \centerline{\includegraphics[width=0.7\textwidth]{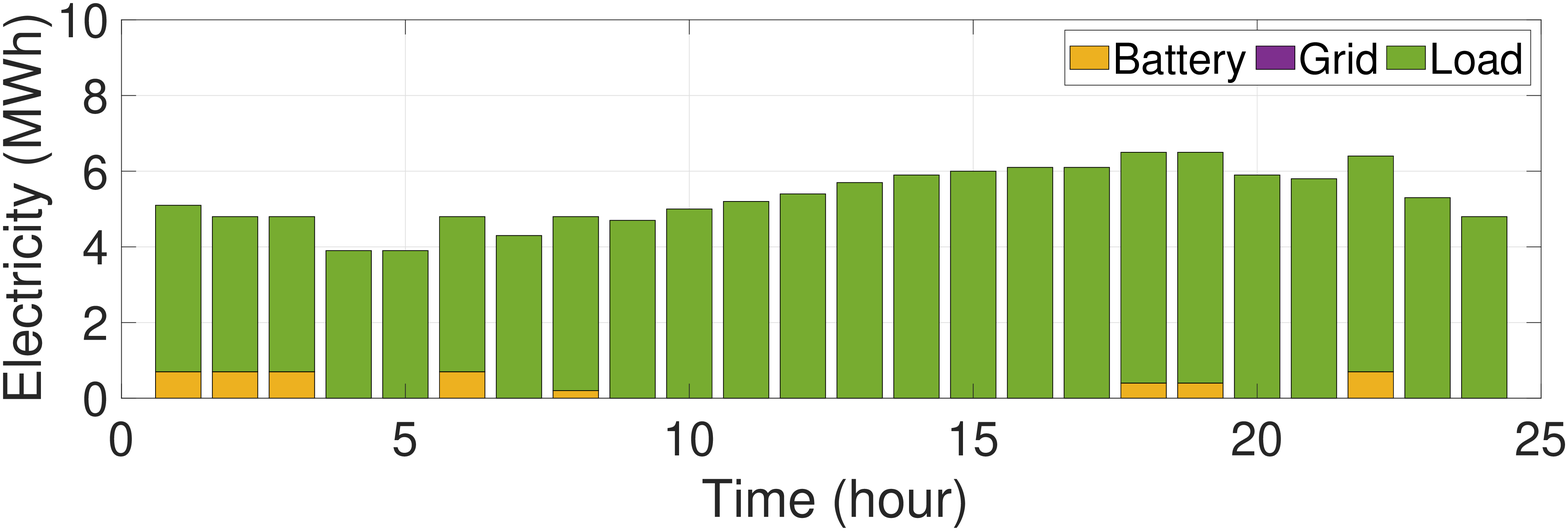}}
  \centerline{\scriptsize{(b) Electricity consumption/charging}}
\end{minipage}
\caption{{Electricity profiles under the proposed algorithm.}}
\label{fig6}
\end{figure}

\begin{figure}
\centering
\begin{minipage}{1\linewidth}
  \centerline{\includegraphics[width=0.7\textwidth]{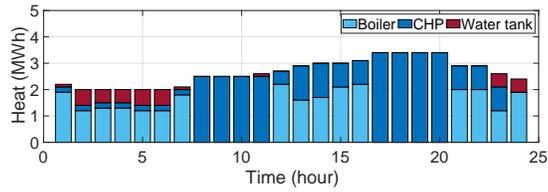}}
  \centerline{\scriptsize{(a) Heat generation/discharging}}
\end{minipage}
\begin{minipage}{1\linewidth}
  \centerline{\includegraphics[width=0.7\textwidth]{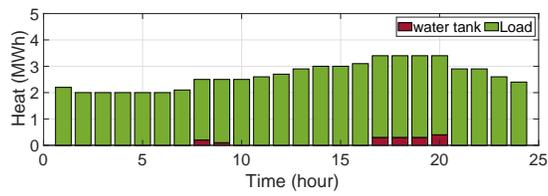}}
  \centerline{\scriptsize{(b) Heat consumption/charging}}
\end{minipage}
\caption{{Heat profiles under the proposed algorithm.}}
\label{fig7}
\end{figure}

\begin{figure}
\centering
\begin{minipage}{1\linewidth}
  \centerline{\includegraphics[width=0.7\textwidth]{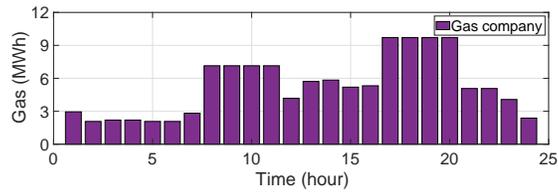}}
  \centerline{\scriptsize{(a) Gas Purchase}}
\end{minipage}
\begin{minipage}{1\linewidth}
  \centerline{\includegraphics[width=0.7\textwidth]{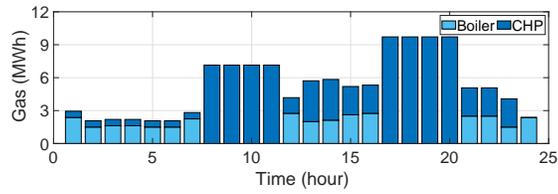}}
  \centerline{\scriptsize{(b) Gas consumption}}
\end{minipage}
\caption{{Gas profiles under the proposed algorithm.}}
\label{fig8}
\end{figure}

Fig.~\ref{fig2}(a) shows that the electricity prices are high during 8:00-11:00 and 17:00-20:00. At this time, the industrial park uses the CHP unit to generate electricity instead of purchasing high-price electricity and simultaneously generate heat to supply the heat demand, which is shown in Figs. \ref{fig6}-\ref{fig8}. At the same time, since the CHP unit generates enough heat, the boiler does not generate heat. 
{At other time, the boiler serves unsatisfied heat demand by consuming natural gas, which is shown in Figs. \ref{fig7}(a) and \ref{fig8}(b). The battery is charged at low-price time 1:00-3:00 and 6:00, and discharged at high-prices time 9:00-11:00 and 20:00, which is shown in Fig. \ref{fig6}. Considering that battery charging and discharging have a small impact on the total cost, the strategy of battery agent is not good enough due to battery capacity limitations and changing electricity prices.  The thermal energy of the hot water tank is charged during 8:00-9:00 and 17:00-20:00 since CHP unit generates extra heat as shown in Figs. \ref{fig7}(b), and the thermal energy of the hot water tank is discharged to supply the heat demand at other time as shown in Figs. \ref{fig7}(a). }
As shown in Figs. \ref{fig6}-\ref{fig8}, the proposed algorithm achieves the multi-energy complementation, multi-device cooperation and multi-energy supply. Therefore, the proposed algorithm can effectively reduce the costs of factories.

\section{Conclusion}
In this paper, we study the multi-energy management problem of an industrial park, which is imperative for today's industrial production. We present a MEMF to achieve centralized training and decentralized execution for energy devices.  To obtain the optimal scheduling policy of each energy device, we design a novel multi-agent deep reinforcement learning algorithm based on the counterfactual baseline and SAC. 
Then, an attention mechanism is introduced to focus on key information, which improves the exploration efficiency of soft actor-critic. At last, based on real data, the performance of DRL algorithms in terms of the number of agents and attention degree of information is analyzed, showing that the proposed algorithm is an effective solution for industrial energy management problems.

 {In this paper, the energy of industrial production is sufficient supplied by energy hubs. In actual industrial parks, some factories may have huge energy demand which cannot be satisfied due to the capacity of transformers. In our future work, the energy scheduling of actual industrial production with insufficient energy supply will be further investigated. Furthermore, the joint optimization of production scheduling and multi-energy generation/utilization is of interest.}
\bibliographystyle{elsarticle-num}
\bibliography{elsarticle-template}

\begin{thebibliography}{}
\expandafter\ifx\csname url\endcsname\relax
  \def\url#1{\texttt{#1}}\fi
\expandafter\ifx\csname urlprefix\endcsname\relax\def\urlprefix{URL }\fi
\expandafter\ifx\csname href\endcsname\relax
  \def\href#1#2{#2} \def\path#1{#1}\fi


\bibitem{Lu2020A}
X. Lu, Z. Liu, L. Ma et al., "A robust optimization approach for optimal load dispatch of community energy hub," \emph{Applied Energy}, vol. 259:114195, Feb. 2020.

\bibitem{Jadidbonab2020Short}
M. Jadidbonab, B. Mohammadi-Ivatloo, M. Marzband and P. Siano, "Short-Term Self-Scheduling of Virtual Energy Hub Plant Within Thermal Energy Market," \emph{IEEE Transactions on Industrial Electronics}, vol. 68, no. 4, pp. 3124-3136, Apr. 2021.


\bibitem{Heidari2020Stochastic}
Heidari A, "Stochastic effects of ice storage on improvement of an energy hub optimal operation including demand response and renewable energies," \emph{Applied Energy}, vol. 261:114393, Mar. 2020.


\bibitem{Liu2020Heat}
N. Liu, L. Zhou, C. Wang, X. Yu and X. Ma, "Heat-Electricity Coupled Peak Load Shifting for Multi-Energy Industrial Parks: A Stackelberg Game Approach," \emph{IEEE Transactions on Sustainable Energy}, vol. 11, no. 3, pp. 1858-1869, Jul. 2020.

\bibitem{Wei2021Distribution}
J. Wei, Y. Zhang, J. Wang and L. Wu, "Distribution LMP-based Demand Management in Industrial Park via a Bi-level Programming Approach," \emph{IEEE Transactions on Sustainable Energy}, vol. 12, no. 3, pp. 1695-1706, Feb. 2021.


\bibitem{Liu2018Co}
Z.~Liu, M.~Adams, R.~Cote, Y.~Geng, J.~Ren, Q.~Chen et al., "Co-benefits accounting for the implementation of eco-industrial development strategies in the scale of industrial park based on emergy analysis," \emph{Renewable and Sustainable Energy Reviews}, vol. 81, no.1, pp. 1522-1529, Jan. 2018.


\bibitem{Shahidehpour2021Two}
M. Shahidehpour, C. Li, X. Wang, W. Huang and T. Nengling, "Two-Stage Full-Data Processing for Microgrid Planning With High Penetrations of Renewable Energy Sources," \emph{IEEE Transactions on Sustainable Energy}, vol. 12, no.4, pp. 2042-2052, May 2018.

\bibitem{Zeng2019Dynamic}
P. Zeng, H. Li, H. He and S. Li, "Dynamic Energy Management of a Microgrid Using Approximate Dynamic Programming and Deep Recurrent Neural Network Learning," \emph{IEEE Transactions on Smart Grid}, vol. 10, no. 4, pp. 4435-4445, Jul. 2019.


\bibitem{Gupta2019Optimal}
S. Gupta, V. Kekatos and W. Saad, "Optimal Real-Time Coordination of Energy Storage Units As a Voltage-Constrained Game," \emph{IEEE Transactions on Smart Grid}, vol. 10, no. 4, pp. 3883-3894, July 2019.

\bibitem{Li2019Energy}
{ Y. Li, H. He, A. Khajepour, H. Wang, J. Peng, "Energy management for a power-split hybrid electric bus via deep reinforcement learning with terrain information". \emph{Applied Energy}, vol. 255:113762, Aug 2019.}

\bibitem{Kou2020Safe}
{ P. Kou, D. Liang, C. Wang, Z. Wu and L. Gao, "Safe deep reinforcement learning-based constrained optimal control scheme for active distribution networks", \emph{Appl Energy}, vol. 264:114772, April 2020.}

\bibitem{Xu2021Multi}
{ Z. Xu, G. Han, L. Liu, M. Martinez-Garcia and Z. Wang, "Multi-Energy Scheduling of an Industrial Integrated Energy System by Reinforcement Learning-Based Differential Evolution," \emph{IEEE Transactions on Green Communications and Networking}, vol. 5, no. 3, pp. 1077-1090, Sept. 2021.}


\bibitem{Ebell2018Reinforcement}
{ N. Ebell, F. Heinrich, J. Schlund and M. Pruckner, "Reinforcement Learning Control Algorithm for a PV-Battery-System Providing Frequency Containment Reserve Power," \emph{2018 IEEE International Conference on Communications, Control, and Computing Technologies for Smart Grids (SmartGridComm)}, 2018, pp. 1-6.}

\bibitem{Mocanu2019On}
E. Mocanu et al., "On-Line Building Energy Optimization Using Deep Reinforcement Learning," \emph{IEEE Transactions on Smart Grid}, vol. 10, no. 4, pp. 3698-3708, Jul. 2019.

\bibitem{Ye2020Local}
Y. Ye, D. Qiu, X. Wu, G. Strbac and J. Ward, "Model-Free Real-Time Autonomous Control for a Residential Multi-Energy System Using Deep Reinforcement Learning," \emph{IEEE Transactions on Smart Grid}, vol. 11, no. 4, pp. 3068-3082, Jul. 2020.

\bibitem{Guo2022Real}
{C. Guo, X. Wang, Y. Zheng, and F. Zhang, ''Real-time optimal energy management of microgrid with uncertainties based on deep reinforcement learning,'' \emph{Energy}, vol. 238:121873, Jan. 2022.}

\bibitem{Sheikhi2016Demand}
A. Sheikhi, M. Rayati, A. M. Ranjbar, "Demand side management for a residential customer in multi-energy systems," \emph{Sustainable Cities and Society}, pp. 63-77, 2016.

\bibitem{Weigold2021Method}
{M. Weigold, H. Ranzau, A. Schaumann, T. Kohne, N. Panten and E. Abele, ''Method for the application of deep reinforcement learning for optimised control of industrial energy supply systems by the example of a central cooling system'', \emph{CIRP Annals}, vol. 70, no. 1, pp.17-20, 2021.}

\bibitem{Wang2021Surrogate}
{X. Wang, Y. Liu, J. Zhao, C. Liu, J. Liu and J. Yan, ''Surrogate model enabled deep reinforcement learning for hybrid energy community operation'', \emph{Applied Energy}, vol. 289:116722, May 2021.}

\bibitem{Ebell2018Coordinated}
{N. Ebell and M. Pruckner, "Coordinated Multi-Agent Reinforcement Learning for Swarm Battery Control," \emph{2018 IEEE Canadian Conference on Electrical and Computer Engineering (CCECE)}, 2018, pp. 1-4.}

\bibitem{Roesch2019Industrial}
{M. Roesch, C. Linder, C. Bruckdorfer, A. Hohmann and G. Reinhart, "Industrial Load Management using Multi-Agent Reinforcement Learning for Rescheduling," \emph{2019 Second International Conference on Artificial Intelligence for Industries (AI4I)}, 2019, pp. 99-102.}

\bibitem{Ebell2019Sharing}
{N. Ebell, M. Gutlein and M. Pruckner, "Sharing of Energy Among Cooperative Households Using Distributed Multi-Agent Reinforcement Learning," \emph{2019 IEEE PES Innovative Smart Grid Technologies Europe (ISGT-Europe)}, 2019, pp. 1-5.}

\bibitem{Busoniu2008A}
L. Busoniu, R. Babuska and B. De Schutter, "A Comprehensive Survey of Multiagent Reinforcement Learning," \emph{IEEE Transactions on Systems, Man, and Cybernetics, Part C (Applications and Reviews)}, vol. 38, no. 2, pp. 156-172, March 2008.

\bibitem{Perera2021Applications}
{A. T. D. Perera and P. Kamalarubann, "Applications of reinforcement learning in energy systems", \emph{Renewable and Sustainable Energy Reviews}, vol. 137:110618, March 2021.}

 
 \bibitem{Nguyen2020Deep}
 T. T. Nguyen, N. D. Nguyen and S. Nahavandi, "Deep Reinforcement Learning for Multiagent Systems: A Review of Challenges, Solutions, and Applications," \emph{IEEE Transactions on Cybernetics}, vol. 50, no. 9, pp. 3826-3839, Sept. 2020.
 
  \bibitem{Lu2020Multi}
{R. Lu, Y. Li, Y. Li and Y. Ding, "Multi-agent deep reinforcement learning based demand response for discrete manufacturing systems energy management", \emph{Applied Energy}, vol. 276:115473, Oct. 2020.}

  \bibitem{Li2021A}
{J. Li, T. Yu, and B. Yang, "A data-driven output voltage control of solid oxide fuel cell using multi-agent deep reinforcement learning", \emph{Applied Energy}, vol. 304:117541, Dec. 2021.}

\bibitem{Sutton1998Reinforcement}
R. S. Sutton and A. G. Barto, "Reinforcement Learning: An Introduction," in IEEE Transactions on Neural Networks, vol. 9, no. 5, pp. 1054-1054, Sept. 1998.

\bibitem{Puterman2009Markov}
Laurence A. Baxter, "Markov Decision Processes: Discrete Stochastic Dynamic Programming," Technometrics, vol. 37, no.3, Mar. 2012.

\bibitem{Wu2021Caching}
{X. Wu, X. Li, J. Li, P. C. Ching, V. C. M. Leung and H. V. Poor, "Caching Transient Content for IoT Sensing: Multi-Agent Soft Actor-Critic," \emph{IEEE Transactions on Communications}, vol. 69, no. 9, pp. 5886-5901, Sept. 2021.}

\bibitem{Wang2020Safe}
{W. Wang, N. Yu, Y. Gao and J. Shi, "Safe Off-Policy Deep Reinforcement Learning Algorithm for Volt-VAR Control in Power Distribution Systems," \emph{IEEE Transactions on Smart Grid}, vol. 11, no. 4, pp. 3008-3018, July 2020.}
 
 \bibitem{Sutton2000Policy}
R. S. Sutton, D. A. McAllester, S. P. Singh, and Y. Mansour, "Policy gradient methods for reinforcement learning
with function approximation," in \emph{Advances in Neural
Information Processing Systems}, pp. 1057-1063, 2000.

\bibitem{Haarnoja2018Soft}
T. Haarnoja, A. Zhou, P. Abbeel and S. Levine, "Soft actor-critic: Off-policy maximum entropy deep reinforcement learning with a stochastic actor," in \emph{International Conference on Machine Learning}, vol. 80, pp. 1861-1870, Jul. 2018.

\bibitem{Oh2016Control}
Oh, J., Chockalingam, V., Lee, H. et al., "Control of memory, active perception, and action in minecraft," in \emph{International Conference on Machine Learning}, pp. 2790-2799, 2016.

\bibitem{Kim2016Hadamard}
{J. H.~Kim, K. W.~On, W.~Lim, J. W. Ha, and B. T. Zhang, 
  ``Hadamard Product for Low-rank Bilinear Pooling,'' 
  2016, arXiv prepint, arXiv: 1610.04325.}

\bibitem{Vaswani2017Attention}
Vaswani, A., Shazeer, N., Parmar, N., Uszkoreit, J., Jones, L., Gomez, A. N., Kaiser, Ł., and Polosukhin, I., "Attention is all you need," in \emph{Advances in Neural Information Processing Systems}, pp. 6000-6010, 2017.

 

\bibitem{Illinois}
The price provided by the State Grid Jiangsu Electric Power Co., Ltd, \emph{http://www.js.sgcc.com.cn/html}. 

\bibitem{PJM} 
PJM hourly load, \emph{https://dataminer2.pjm.com}.

\bibitem{Institute}
Renewables.ninja, \emph{https://www.renewables.ninja}. 


\bibitem{DDPG}
 T. P. Lillicrap, J. J. Hunt, A. Pritzel et al., "Continuous control with deep reinforcement learning," in \emph{International Conference on Learning Representations}, 2016.
 
 \bibitem{MADDPG}
 R. Lowe, Y. Wu, A. Tamar et al, "Multi-agent actor-critic for mixed cooperative-competitive environments," in \emph{Advances in Neural Information Processing Systems}, pp. 6382-6393, 2017.
 
 
 
 





\end{thebibliography}

\end{document}